# An Effective Approach for Mobile ad hoc Network via I-Watchdog Protocol

Nidhi Lal

*Indian Institute of Information Technology*

*Abstract* — Mobile ad hoc network (MANET) is now days become very famous due to their fixed infrastructure-less quality and dynamic nature. They contain a large number of nodes which are connected and communicated to each other in wireless nature. Mobile ad hoc network is a wireless technology that contains high mobility of nodes and does not depend on the background administrator for central authority, because they do not contain any infrastructure. Nodes of the MANET use radio wave for communication and having limited resources and limited computational power. The Topology of this network is changing very frequently because they are distributed in nature and self-configurable. Due to its wireless nature and lack of any central authority in the background, Mobile ad hoc networks are always vulnerable to some security issues and performance issues. The security imposes a huge impact on the performance of any network. Some of the security issues are black hole attack, flooding, wormhole attack etc. In this paper, we will discuss issues regarding low performance of Watchdog protocol used in the MANET and proposed an improved Watchdog mechanism, which is called by I-Watchdog protocol that overcomes the limitations of Watchdog protocol and gives high performance in terms of throughput, delay.

## I. Introduction

IN the mid of 1990's, Mobile ad hoc network became very famous topic in the research area of networking. Mobile ad hoc network is a wireless technology and it does not hold any infrastructure; nodes in the MANET environment are dynamic in character and do not relay on any topology. They are scattered in nature and do not rely on any central authority. In a MANET, [11] each node can take responsibility of a router as well as take a role as a host. The nodes in the mobile ad hoc network are linking and Communicate with each other all the way through the usage of radio waves. [1]MANET supports fast establishing of networks so they encompass very high degree of flexibility, the only necessity is to provide a new set of nodes with some degree of wireless communication range. [17]If the nodes are within the same radio waves wireless communication rang than they can communicate directly otherwise they can communicate with their respective destination node with the help of intermediate nodes.

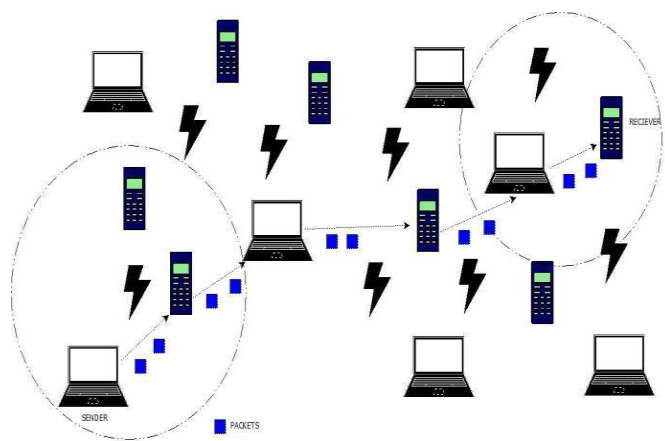

Fig 1. The sketch out of MANET

In the above figure, many numbers of nodes are in the network, in which, one is acting as a sender node and another one is the receiving node. Sender wishes to propel data packet to his subsequent receiving node. For initiating this communication, the sender can send these data packets via the help of intermediate nodes, which are within the communication range of the sender node. By following this strategy, the sender node sends all data packets to the respective receiver node. [2]There are various types of the mobile ad hoc networks, such like, vehicular ad hoc network (VANET) [12] that is used for making the communication between the vehicles, (IMANET) internet based mobile ad hoc networks which is used to link mobile nodes to network gateways and Tactical MANET which is used in the application of military. [3]Meant for routing of the packets among the mobile device nodes a routing protocol is obligatory. [37]The routing protocol should design and chooses in such a way that it provides high reliability, security, power efficient, avoid overhead and provide best quality of service as well as should consider the unidirectional links also. So by taking these points into account, there are various methodologies are proposed like, AODV, DSDV, DSR, CBRP. [4]AODV maintains route on the demand such that traffic of the network remain minimum and it uses distance sequence number for surety of the loop free route. [5] Security threats and packet loss due to transmission error are major challenges of the MANET. [18]Destination-Sequenced Distance-Vector Routing (DSDV) maintains a table for routing and follows the concept of bellman-ford algorithm, it is basically used to prevent the network from the formation of a loop. DSDV make sure that the network does not restrain any cycle or loop. DSDV has certain disadvantages like it is not power saving and does not worth for networks which are highly dynamic. [19]Dynamic Source Routing (DSR) is a source routing protocol which is similar to AODV but it does not contain latest updated information regarding the network therefore it leads to inconsistency in the routing tables. [20]A cluster-based routing protocol (CBRP) is a routing protocol in which, nodes of a network make a group and that group is called a cluster, after forming this cluster they uses a clustering algorithm to determine the cluster head among the nodes in that group. [30-32]A Mobile Location Aware Information System is also proposed for control of the presence of a non-intrusive by use of technologies which are based on the global positional system (GPS) and light weight indoor location system. [33-35]This technology can be used universally and applies very minimum cost. [5]Due to the decentralized environment of the MANET, they are constantly susceptible to black hole attack and the recital of the AODV routing protocol decrease. To triumph over this dilemma, Watchdog protocol with AODV is commenced which builds recognition of malicious nodes.

Watchdog protocol uses local information of the next hop node and overhears it. If it gets that it spending time of the packet is exceeded above the predefined threshold then it marks that node as malicious, this way Watchdog protocol detects malicious node in the network. [6, 10]Watchdog protocol has some disadvantages that it does not find link transmission error due to congestion in the network as well as it does not support high mobility of large number of nodes in the network and give a wrong report about the malicious node which eventually decreases the system throughput and performance. [7]Mobile ad hoc networks are more vulnerable to transmission errors than fixed wired network because of their wireless nature, environmental conditions, network congestion etc. In this paper, an improved Watchdog mechanism is presented which identifies the malicious node in the network as well as spots the network congestion. We give the name of this improved Watchdog protocol as I-Watchdog protocol. The proposed work is implemented in ns2-simulator and gives very high performance in terms of throughput, packet delivery ratio and end-to-end delay. Section-2 will describe the AODV protocol, Section-3 will introduce about some variety of security attacks, Section-4 describes the activity of black hole attack in the network, Section-5 will describe Watchdog protocol, Section-6 will introduce about proposed improved Watchdog protocol (I-Watchdog protocol) and section-7 is containing simulation parameters and Section-8 will contain comparison, results of Watchdog protocol and proposed improved Watchdog protocol.

## II. AODV PROTOCOL

MANET applies Ad hoc On Demand Distance Vector (AODV) routing protocol for transmitting the packet from the source towards their particular target node. [8]AODV each point in time determine route when network wishes it. AODV bring into play route request message (RREQ) for creation the route request from source to target which enclose the distance sequence number. This communication is basically does neighbor discovery and it is broadcasting in nature. If the intermediary neighboring node have a path to the resultant destination, then it propels the route reply message (RREP) reverse to the source. If the intermediary adjacent nodes have no path to the destination then it generates reverse route entry towards the source and broadcast RREQ message to its neighbors. This course of action will be on recurrence until the destination route is not set up. Just the once the route is found to the target node then the RREP message is unicast from the current node to the source node and this RREP message include destination sequence number and hope count. Later than receiving the RREP message, source node brings up to date its routing table only in the provision when coming destination sequence number is bigger than the prior. Upon receiving on the several RREP, source opt for greater sequence number with smallest hop count and renew its routing table information. [22]In the AODV routing protocol, each adjacent node in the network background maintains track information about the status of the link by keeping an eye on the link and when there is found the splintering in link of the route then the RERR message is promoted by this node which detects the link breakage, to all the nodes in the network to broadcast this link status information. Figure 2 demonstrates the appearance of AODV routing protocol. Here, node marked by S represents a source node and node marked by D work as destination node.

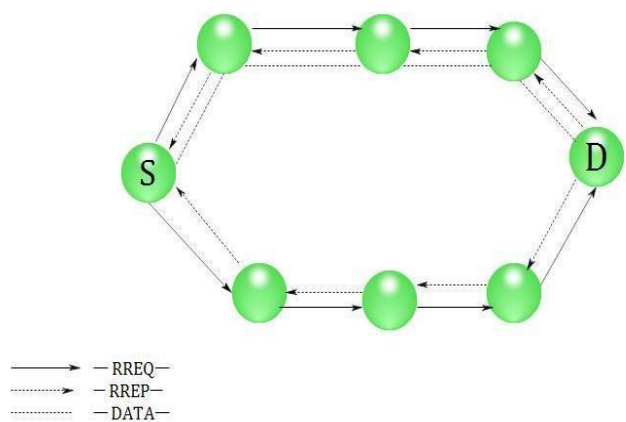

Fig: 2. AODV routing protocol method

## III. VARIETY OF SECURITY ATTACKS

There is a range of attacks probable in MANET. [31] Because of the absence of any central authority and dynamic and distributed nature, mobile- ad hoc network environment can have a lot of security breaches and problems. Security does direct impact on the performance of the network because when the network is more secure then always there is a high possibility of successful transmission of packet from the sender to respective receiver. This will directly lead to high throughput and minimizes the end-to-end delay. But due to MANET is wireless in nature, there is always the possibility of attacks on the network. [21]A few of the security attacks are characterized below:

- BLACK HOLE ATTACK: In this attack, every packet which is sent out from the source towards the relevant target node, is dropping by black hole node. This attack is illustrated in detail in the section- IV
- PARTITION OF THE NETWORK: In this attack, a path is always present from the sender node to the respective receiver node, but nodes cannot communicate with each other.
- DENIAL OF SERVICE: In this attack, there is constraint on the nodes of a network for conveyance the packet as well as in receipt of the packet.
- SLEEP DEPRIVATION: [28, 36] In this attack, the nodes are forced to be sleeping, means to force for use its battery power.
- INFORMATION THEFT: [36] In this attack, the whole information which is reside inside the packet is read by unauthorized entity.
- INTRUSION: [36] In this attack, an unapproved person can have right to use services and those services are constrained to that entity.
- TAMPERING: [36] In this attack, Data is modified by an unauthorized entity.
- WORM HOLE ATTACK: [37] In this attack, basically the attacker node compromised with any host in the network and record the packet at one point of location and after tunneling to another location it again sends back to the network.

## IV. BLACK HOLE ATTACK

MANET does not encompass any central authority and there is a lack of infrastructure so that is defenseless to black hole attack. The black hole attack is an attack on a network who hurriedly dwindle network performance by dropping the packets. [16]When a black hole node (malicious node) present in the network, it always advertises itself with the highest sequence number and minimum hop count. [23]Black hole node always tries to attract and capture the attention of the source node by ensuring them that it has the shortest path towards the destination node. The black hole attack is very dangerous in the network environment and it leads to the system to a denial of service (DOS) attack. When it obtains RREQ from the source node, it instantaneously propels RREP respond to the source enclose very large sequence number and lowest hop count. This is the nature of the black hole node that it tries to get attention in the network from source point of view, [17] that's why it advertises itself with very high sequence numbers. Upon receiving such RREP from this black hole node, source node thinks that it has valid fresh route to the destination because it have a high sequence number with minimum hope count and starts forwarding of data packets towards this black hole node. Upon receiving data packets, the black hole node drops all data packets and system performance degrades rapidly. Figure 3 shows the activity of black hole attack, in this figure red circles is used for denoting malicious node and green circle to represent valid nodes.

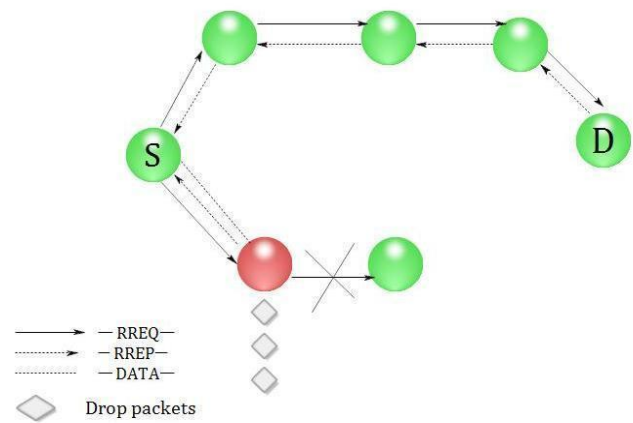

Fig: 3. The course of action of black hole attack

On top of figure, source node S launch RREQ to its neighbor node. The malicious node does not further send RREQ to its neighborhood node and propel RREP to source node S with high sequence number and least hope count. After getting such RREP, the source node sends data to this malicious node and malicious node drop all packets and system is in beneath the black hole attack.

## V. WATCHDOG PROTOCOL

To avoid the problem and detection of this black hole attack, [9] Watchdog protocol is introduce. In this protocol, every node is work as an observer to watch the working of its next hope neighborhood node. It collects transmission information of this node and observes that node correctly forward to its next hope neighborhood node along with the correct destination route. [13]This protocol measures the sending time of the next hope node. If the sending time of the next hop neighbor is greater than the packet storing time and exceeds above some defined threshold of the network, then Watchdog knows that system is under black hole attack and it immediately mark this node as a malicious node. The Watchdog protocol announces the existence of the malicious node in the network by generating the alerts. The benefit of the Watchdog protocol is that, they make use of only local information and are proficient to spot the malicious node.

They can resolve the predicament of black hole attack which demonstrate the way to denial of service attack (DOS) in MANET network. [14]Watchdog protocol act as a very good intrusion detection system mechanism in the network. However, [15] there are certain disadvantages regarding to this protocol such that it decreases the network performance in terms of throughput, it does not support mobility with high number of nodes, and it doesn't detect the actual reason of the packet loss. To overcome these disadvantages of this Watchdog protocol, the improved Watchdog mechanism is proposed which perfectly distinguishes the packet loss due to congestion or due to the presence of a malicious node in the network. Our improved Watchdog protocol also supports a high degree of the mobility and enhances the performance.

## VI. PROPOSED WORK

In this paper, the improved Watchdog protocol is proposed with some modifications to overcome the problem related to the early Watchdog protocol. This improved Watchdog protocol is very efficient to detect the actual reason for the packet loss. Because MANET is a wireless technology, mobile node devices of the MANET are independent to move anywhere so the mobility is very high and Watchdog protocol does not support a high degree of mobility but our I-Watchdog protocol supports a very large number of nodes with a high degree of mobility. Also, due to the wireless nature of the mobile ad hoc networks, they are more vulnerable to congestion and Watchdog protocol does not detect the network congestion and link error of transmission. It just observes that whenever the sending time of packet greater than the packet storing time, it sends alert in the system and marks the node as malicious. Watchdog protocol does not find the actual cause of packet loss and this leads to low throughput and system performance degrades. But proposed I-Watchdog protocol does not take a decision about the node very easily, because the packet loss also happens due to network congestion, it implements some modifications to the existing protocol. I-Watchdog protocol gives good results in throughput, packet delivery ratio and end-to-end delay as compared to Watchdog protocol. In the below figure shows the algorithm that we will use in the implementation of our I-Watchdog protocol.

ALGORITHM OF I-WATCHDOG PROTOCOL:

1. If (sending time of packet > Packet storing time ) else go to step 8
2. Calculate d = sequence no of suspected node – sequence no of current node
3. If d is very large and within the range of suspected node's sequence number then go to step 4 else go to stop 7.
4. Then calculate % packet loss of suspected node to be malicious
5. If (% of packet loss > threshold of % packet loss)
6. Then Mark the suspected node as malicious
7. Else call local repair of link function
8. stop

In the above algorithm, Watchdog observes the next hope node activity. Whenever it detects that node's sending time exceeds the packet storing time then it does not directly mark node as malicious, it further checks for the sequence number. It calculates the difference which is denoted by d in the above algorithm, between the sequence number of suspected nodes and the sequence number of itself. If this difference d is very close to the suspected sequence number and it is very far from the sequence number by itself then it checks for the packet loss of percentage. For example, consider the below 2 cases:

- CASE 1:

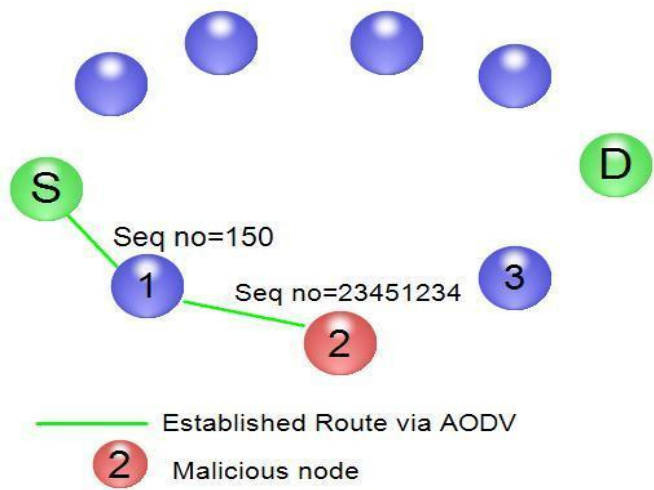

Fig 4. CASE 1

From the above figure 4, suppose node 1 act as a Watchdog and observes the next hop node that is node 2 and node 2 is suspected to be malicious because its sending time is exceeded over packet storing time. Then node 1 calculates difference d. Suppose node 2 is a malicious node so it contain a very large sequence number for example the sequence number of node 2 is 23451234 and sequence number of node 1 is 150 then difference d is evaluated to be d = (23451234-150) =>23451084, which is very close to the sequence number of node2 (suspected malicious nodes) and very far from node 1. After calculating this value of d, to ensure the suspected node is malicious it checks for the percentage of packet loss. If this percentage of packet loss has exceeded the predefined threshold value of percentage of packet loss then I-Watchdog protocol mark this node as malicious and send alert in the network about the malicious node. If suspected malicious node's percentage of packet loss is less than the threshold value than it does local repair link because it indicates that packet loss is due to the network congestion, transmission errors and the suspected malicious node is not malicious.

- CASE 2:

In case 2, when difference d is not close to the suspected malicious node as well as node which act as Watchdog, then there is a confirmation that node is not malicious and sending time exceeds the threshold due to transmission errors and congestion so local repair of link function is called. For example node 1 sequence number is 150 and sequence numbers of suspected malicious nodes is 170 then the differences will be d = (170-150) =>20, which is not close to 170 as well as 150. In this case, local repair of link function can be directly called. This algorithm of the I-Watchdog protocol gives better performance and high throughput.

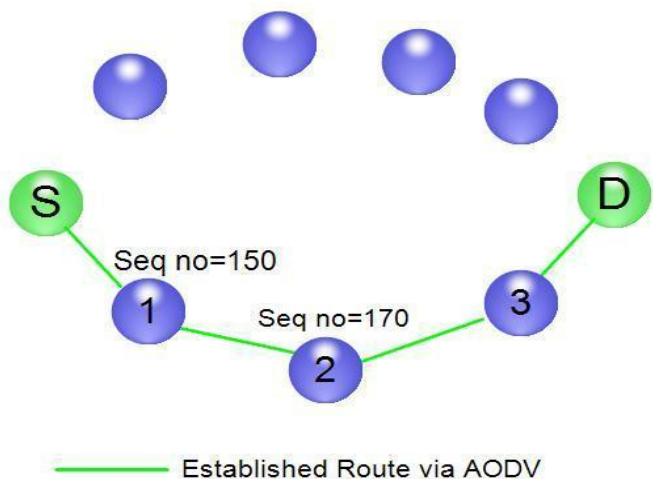

Fig 5. CASE 2

In case 2, when difference d is not close to the suspected malicious node as well as node which act as Watchdog, then there is a confirmation that node is not malicious and sending time exceeds the threshold due to transmission errors and congestion so local repair of link function is called. For example node 1 sequence number is 150 and sequence numbers of suspected malicious nodes is 170 then the differences will be d = (170-150) =>20, which is not close to 170 as well as 150. In this case, local repair of link function can be directly called. This algorithm of the I-Watchdog protocol gives better performance and high throughput.

Now for better understanding of implemented I-Watchdog protocol algorithm following figure shows a flowchart. This flowchart well describes about the mechanism of the improved watchdog protocol. It first calculates the difference d and then check this d is close to the suspected malicious sequence number or not. If it is in the range, then again check for the percentage of packet loss. If it is greater than the predefined threshold then it provides surety that the node is malicious. If d is not within the range of malicious node sequence number then it indicated that sending time exceeds the packet storing time due to the presence of congestion in the network and local repair of link function is directly called.

shown in table 1 which we will use in the simulation of Watchdog protocol and I-Watchdog protocol.

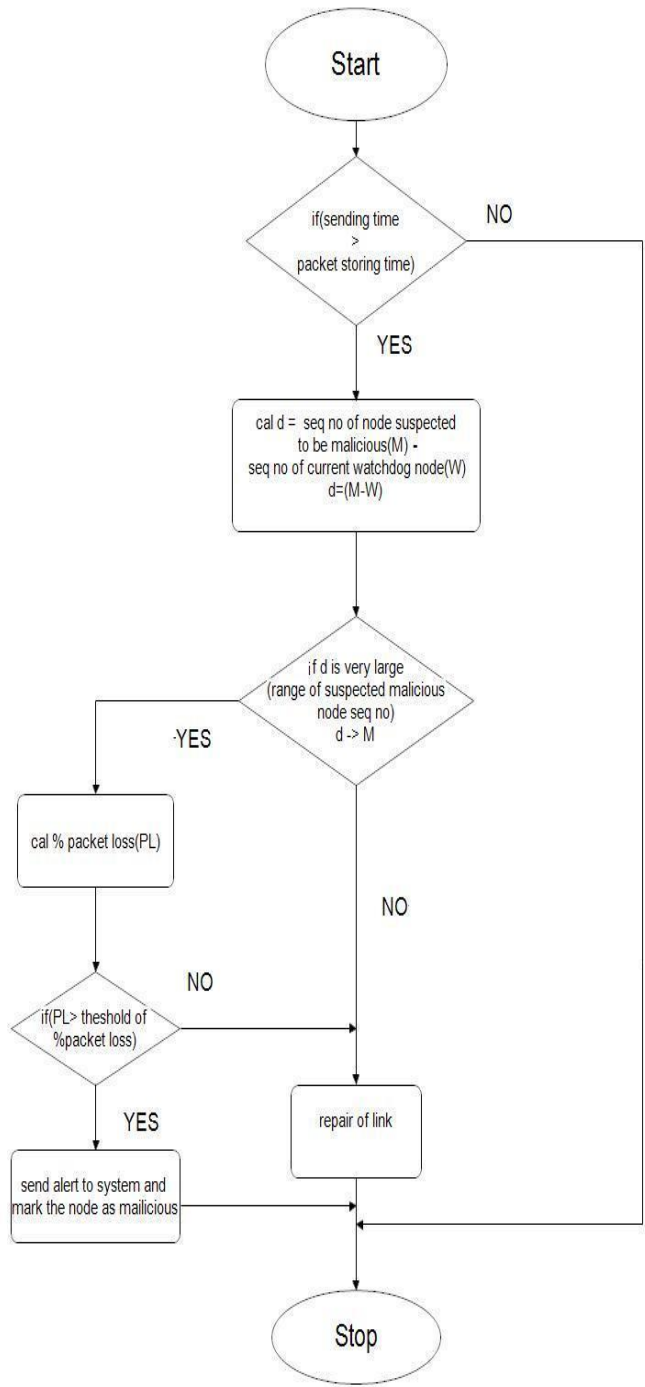

Fig 6. Flowchart of I-Watchdog protocol algorithm.

## VII. SIMULATION

I-Watchdog protocol is implemented in network simulator (NS-2) in Ubuntu platform. In this paper, we are comparing the performance of I-Watchdog protocol with existing Watchdog protocol in terms of throughput, packet delivery ratio and end-to-end delay. The simulation parameters are

## VIII. RESULTS AND COMPARISION

In this paper, I-Watchdog protocol implemented as an improved Watchdog protocol for mobile ad hoc networks. I-Watchdog protocol gives better results in terms of throughput, packet delivery ratio and end-to-end delay. Further, we will equate the performance of both the protocols by in terms of these attribute by plotting the X-graph in NS-2.

TABLE I
SIMULATION PARAMETERS FOR WATCHDOG

| Sr no. | Parameter | Value |
|---|---|---|
| 1 | Simulator | NS-2 |
| 2 | Channel type | Channel/Wireless channel |
| 3 | Radio Propagation Model | Propagation/ Two ray ground wave |
| 4 | Network interface type | Phy/WirelessPhy |
| 6 | MAC Type | Mac /802.11 |
| 7 | Interface queue Type | Queue/Drop Tail |
| 8 | Routing procedure(protocol) | AODV |
| 9 | Antenna | Antenna/Omni Antenna |
| 10 | Type of traffic | CBR |
| 11 | Area ( M*M) | 500 * 500 |
| 12 | Simulation Time | 250 sec |
| 13 | No of Nodes | 50 |

A. *THROUGHPUT*

The principal performance is measured in the relations of the throughput. Throughput is represented in bits per second (bps) and it is the number of packets which is received in per unit of time. Figure 7 represents the throughput of the Watchdog protocol.

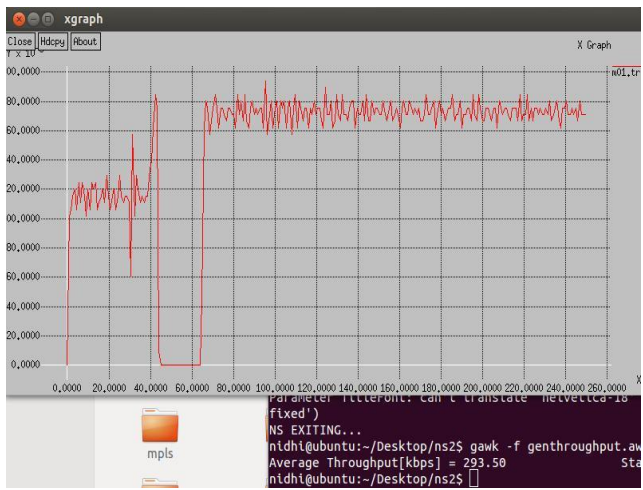

Fig 7. Throughput of the Watchdog protocol

In the above figure, shows graph of the throughput corresponding Watchdog with AODV protocol. The figure represents the average throughput of the system using Watchdog protocol is 293.50 Kbps.

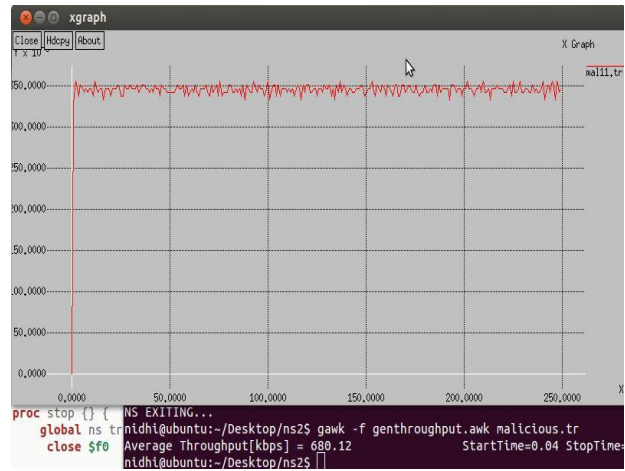

Fig 8. Throughput of I-Watchdog protocol with AODV

In the above figure, the graph for the throughput of I-Watchdog protocol is shown. It is clear from the graph is, the average throughput calculated for me-Watchdog protocol is very high and comes out to be 680.12 kbps, which is very high as compared to the existing Watchdog protocol.

B. *PACKET DELIVERY RATIO/FRACTION*

Packet delivery ratio (PDR) is the portion with reference to the data Packets received by the target node to folks propel by the source node. This evaluates the ability of the protocol performance and its efficiency.

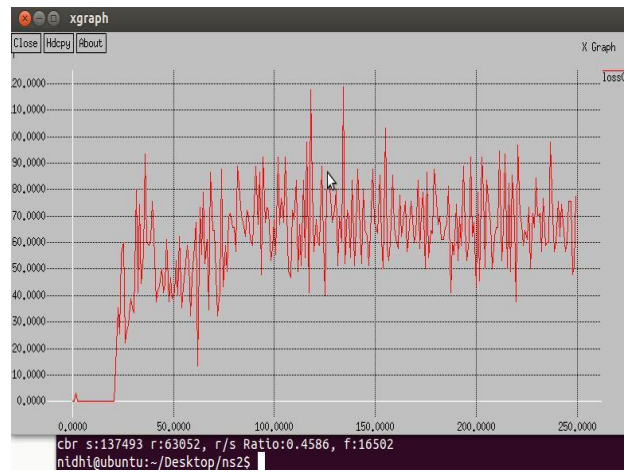

Fig 9. Packet delivery ratio of Watchdog protocol

It is shown from the above figure that the PDR of the system which uses the Watchdog protocol is less. The difference between the sending and receiving packet is 74441 means that 74441 packets are not received by the destination so that we can say the 74441 packets are dropped.

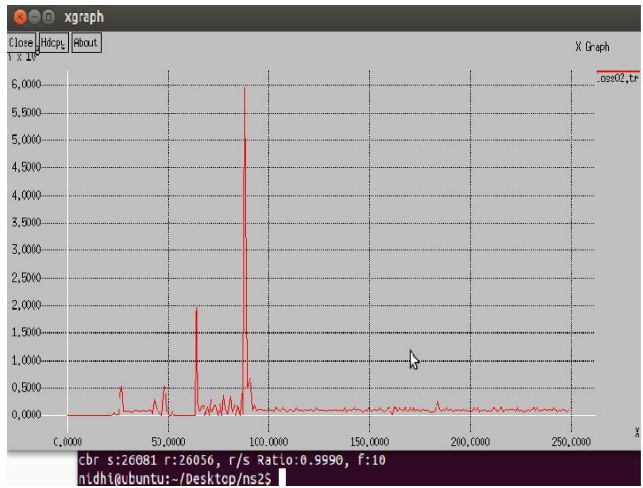

Fig 10. Packet delivery ratio of the system within Watchdog protocol

The figure shows clearly that the numbers of dropped packets are very few. The difference between the number of sending packets and the number of receiving packets is about 25, which is very less as compared to the Watchdog protocol. So that it is clear that the I-Watchdog protocol gives better results and performance than the existing Watchdog protocol.

C: *END-TO-END DELAY*

End to end delay is the quantity of the time which is occupied to sending packets from source to their respective destination to receive those packets. More delay can lead to low performance of the MANET and low delay is the indication of high efficiency and speed of the network.

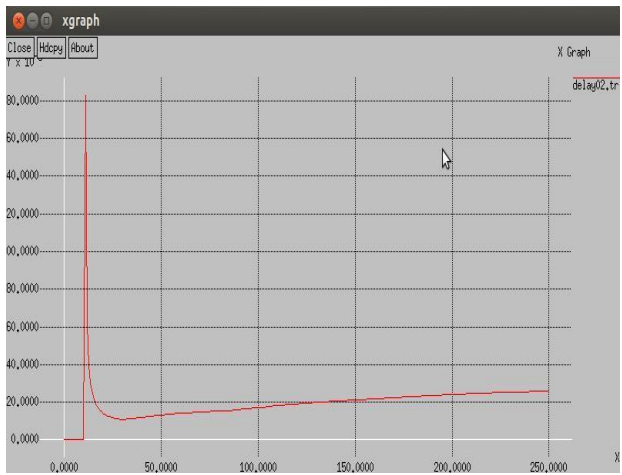

Fig 11. End-to-end delay of the network with Watchdog protocol

The above figure shows end-to-end delay of the network, and it is clear from the graph that it comes out to be about 80 ms which is very high and can highly degrade the system performance. It is fundamentally the total time, which is occupied by the network to send the all packets from source to destination. Here the delay is 80 ms represent the total time is 80ms to send packets from the source node to the destination mobile node in the MANET.

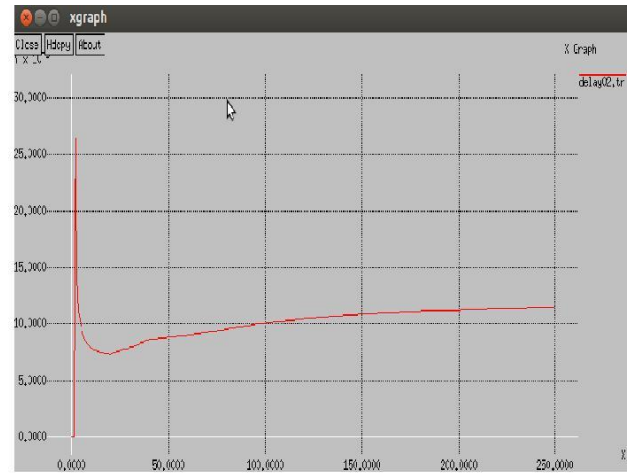

Fig 12. End-to end delay of the system using I-Watchdog protocol

The above figure shows graph of the end–to-end delay of the network which is time to sending the packets from source mobile node to the destination mobile node. The end-to-end delay is about 30ms which is represented by the above X-graph which is very low as compare to the existing Watchdog protocol. So that we can come into the conclusion that proposed I-Watchdog protocol [24-27] requires very less time to send packets from source to destination, it also drops very few packets because of PDR is high for I-Watchdog protocol as compared to existing Watchdog protocol as well as it gives very high throughput which enhances the system performance.

IX. CONCLUSION

MANET is a wireless ad hoc network which is infrastructure less, dynamic and distributed in nature. The attacks are the key sanctuary encounter of MANETs. However I-Watchdog protocol provides a way that can enhance the system performance and detects the congestion, transmission error, link error in the network. It gives high performance and supports a very high number of nodes and provides minimum delay with enhanced throughput. I-Watchdog protocol overcomes the limitation of the previous Watchdog protocol, results less number of dropped packets, high throughput and minimum delay. It can easily detect that the delay and packet drop event is occurring due to transmission error or any attack in the network. If it is due to attack, then an alert is generated by Watchdog node and broadcast information regarding malicious node in the entire network. If it is due to any transmission error then local repair of link function is called. I-Watchdog protocol supports very high degree of the mobility and also supports the dynamic and distributed nature of the MANET. The future work will be on the prevention of the packets from being alternation by the malicious node in the network, in such a way the main focus will be on the integrity and confidentiality of the contents inside the packet. For further enhancing security and efficiency, we provide some authentication techniques and repairing of link methods such that reliable delivery of packets from source to destination will take place.

ACKNOWLEDGMENT

The author would like to be grateful for Mr. Vijay Bhasker Semwal working as a PH.D Scholar in IIIT-Allahabad for the great support and all the contemporaries and fellows, for precious involvement in this paper.

**Nidhi Lal,** is a M.tech student in the Area of Wireless Communication and Computing at Indian Institute of Information Technology, Allahabad. She received her graduate degree from Galgotias college of Engineering and Technology.